\documentclass[a4paper,10pt]{article}
\usepackage[utf8]{inputenc}
\usepackage{amsmath,amssymb,amsfonts}
\usepackage{algorithmic}
\usepackage{graphicx}
\usepackage{multirow}
\usepackage{algorithm}
\usepackage{multicol}
\usepackage{float}
\usepackage{caption}
\usepackage{subfig}
\usepackage{adjustbox}
\usepackage{comment}
\usepackage{textcomp}
\usepackage{geometry}
\usepackage{url}

	\title{Window-based Streaming Graph Partitioning Algorithm}
\author{{Md Anwarul Kaium Patwary}, {Saurabh Garg, and Byeong Kang}}
\date{}

\begin{document}
	\maketitle
	\begin{abstract}
		In the recent years, the scale of graph datasets has increased to such a degree that a single machine is not capable of efficiently processing large graphs. Thereby, efficient graph partitioning is necessary for those large graph applications. Traditional graph partitioning generally loads the whole graph data into the memory before performing partitioning; this is not only a time consuming task but it also creates memory bottlenecks. These issues of memory limitation and enormous time complexity can be resolved using stream-based graph partitioning. A streaming graph partitioning algorithm reads vertices once and assigns that vertex to a partition accordingly. This is also called an one-pass algorithm. This paper proposes an efficient window-based streaming graph partitioning algorithm called WStream. The WStream algorithm is an edge-cut partitioning algorithm, which distributes a vertex among the partitions. Our results suggest that the WStream algorithm is able to partition large graph data efficiently while keeping the load balanced across different partitions, and communication to a minimum. Evaluation results with real workloads also prove the effectiveness of our proposed algorithm, and it achieves a significant reduction in load imbalance and edge-cut with different ranges of dataset.
	\end{abstract}
\section{Introduction}	
The scale of graph data is becoming larger and the trend will continue to grow rapidly with the emergence of different applications (e.g. web-graph, social networks, road networks and biological networks) that deal with massive inter-connectivity\cite{7926432}. Consequently, most real-world applications require distributed computation due to the emergence of these large graphs. To complete the distributed computation of any application, we need to partition an entire graph across machines in a cluster for faster localized processing. This is a vital process in distributing the load.

Graph partitioning cuts a graph into several disjoint sub-graphs with the aim of minimizing the edges between these sub-graphs while retaining almost the same number of vertices in every partition. Imbalance amongst computational load in a distributed environment produces inefficient applications. Besides minimization of communication, the load balancing also must be considered in graph partitioning. These two aspects make graph partitioning an essential pre-processing task for efficient computational speed of different real-world graph applications.

In traditional graph partitioning\cite{Malewicz:2010:PSL:1807167.1807184,4536261}, the entire graph must be loaded into memory for partitioning and processing. This potentially requires huge storage/memory capacities due to the large-scale of the data, which in some cases may increase continuously over time. Consequently, the traditional partitioning algorithm requires high computational costs and memory for the partitioning task. In addition, this also affects the graph data processing for different applications (e.g. PageRank, Shortest Path). These are the main motivations for this study, that proposes a streaming graph partitioning algorithm to partition large graphs efficiently. Streaming graph partitioning is a new variant of the graph partitioning problem, which aims to deal with time-evolving graph datasets. The streaming partitioning technique is also known as a single pass algorithm, as the data can be seen only once in this partitioning algorithm. The streaming graph partitioning algorithm was introduced by Stanton\cite{Stanton:2012:SGP:2339530.2339722}. It aims to provide efficient graph processing by reducing memory bottlenecks by allocating the graph data as it arrives rather than loading the entire graph into memory. It was a welcomed approach, as graph datasets are rapidly growing day by day. As a result, the streaming graph-partitioning is now playing a vital role in overcoming the issues that traditional partitioning can not.

Several studies have been conducted on streaming graph partitioning \cite{Stanton:2012:SGP:2339530.2339722,Stanton:2014:SBG:2634074.2634169,Ahmed:2013:DLN:2488388.2488393,Tsourakakis:2014:FSG:2556195.2556213,Abdolrashidi7584916}, however, many of the studies use a synthetic dataset to evaluate algorithm performance rather than real-world graphs. The studies also assume that the graphs are already localized on the disk and the stream of data will be in a particular order (e.g. Breadth First Search, Depth First Search). In real-world scenarios, graph data do not come in a certain order. Consequently, there is no scope to use any particular stream ordering in real-world graph application scenarios.

In this paper, we propose a window-based streaming graph partitioning technique to obtain better partitioning performance and to reduce edge-cut whilst keeping load imbalance as low as possible. The key idea of this algorithm is that the window-based stream of graph data has more information on vertex allocation because usual single-pass graph partitioning only uses the presence of a vertex to determine the partition for that vertex. The window-based algorithm does not consider any stream order when receiving the stream of data input.  We argue that this technique improves the partitioning performance with regard to the following aspects: i) It balances computational loads among machines; ii) It addresses scalability, as it accepts any range of datasets; and iii) It reduces the communication between machines (by reducing the number of edge-cuts). The contributions of this paper are follows:

\begin{itemize}
	\item A window-based streaming graph partitioning technique that aim to reduce the number of edge-cuts by maintaining a balanced partition.
	\item A streaming window that helps to obtain more information associated with a vertex before a vertex is assigned to a partition.
	\item An algorithm that checks the number of edges of a buffered vertex in the window, which helps achieve better partitioning performance when assigning the vertex.
	
\end{itemize}

In Section 2, we discuss the related work and the most recent advancements in streaming graph partitioning. Our proposed streaming graph partitioning algorithm is explained in Section 3. In Section 4, the dataset, and evaluation setup are discussed. In Section 5, results for this study are analyzed. Finally, we discuss conclusion of this study in Section 7.

\section{Related Work}

Recently, there has been considerable interest in the design of an algorithm and framework to handle massive graph data in a streaming manner. Steaming graph-data can be partitioned into a cluster of nodes; the graph access pattern could be done via online or offline processing. Streaming graph partitioning is considerably efficient because the graph loader or partitioner does the partitioning task while receiving the graph data in a streaming manner. A near-optimal traditional graph partitioning algorithm called METIS was proposed in the early graph-partitioning era\cite{kernighan_graph-partitioning1970}. METIS is the de facto standard for near-optimal partitioning in distributed graph partitioning. METIS can reduce the communication costs among distributed machines despite having a lengthy processing time for small graphs. However, METIS is not suitable for processing medium or large graph datasets\cite{kernighan_graph-partitioning1970}.

Graph partitioning can be categorized into two types: Vertex-Cut-based and Edge-Cut-based. In Table 1, we summarize and compare the most recent stream-based partitioning algorithms. In the following sections, we provide details of related work.

\subsection{Vertex-cut Partitioning}

A scalable streaming partitioning approach was proposed by Wang and Chiu \cite{Wang6691619} with the aim of achieving a low complexity system. This partitioning technique aim to reduce the number of edges between partitions, and consequently reduces the communication cost of query processing.  A streaming vertex-cut partitioning algorithm, High Degree Replicated First (HDRF), was proposed by Petroni et al. \cite{Petroni:2015:HSP:2806416.2806424} to utilise the vertex characteristics. The study used a Greedy vertex-cut approach, in which the high-degree (number of edges of a vertex) vertices replicate first to minimize and avoid unnecessary vertex replication. This algorithm achieved a significant improvement in stream-based partitioning compared to previous algorithms\cite{Low:2012:DGF:2212351.2212354}. HDRF achieves nearly 2x the speedup than traditional Greedy placement and almost three times faster than using a constrained solution. Sajjad et al. proposed a scalable streaming graph partitioning technique called HoVerCut \cite{7584914}, which provided horizontal and vertical scalability for the graph partitioning system. HoVerCut used multi-threading with a windowing technique to share incoming edges between the threads. However, in that the window that contains the edges does not update over time. This may create performance degradation and it is not suitable for dynamic datasets. 

Real-world graphs, for example, social networks, typically follow a power-law degree distribution. Partitioning power-law graphs is very difficult. PowerGraph\cite{Gonzalez:2012:PDG:2387880.2387883} aims to reduce inter-partition communication by computing edges over vertices of power-law graphs. It follows the GAS (Gather, Apply, and Scatter) model and uses a vertex-cut partitioning technique. It distributes replicas of vertices into multiple machines to parallelize the computation.

Another variant of PowerGraph streaming partitioning was proposed by Xie et al\cite{DBLP:journals/corr/XieLZ15} called S-PowerGraph. S-PowerGraph also used vertex-cut partitioning. This method is suitable for partitioning skewed natural graphs and was found to outperform algorithms in previous studies with regard to the acceptable imbalance factor.  

\subsection{Edge-cut Partitioning}

STINGER\cite{Riedy:2013:MSD:2425676.2425689}, a framework to analyze streaming graph structured data, was proposed to facilitate portability, productivity, and performance for the research and development of big data. Its motivation was based on contemporary issues, which can be formulated on the basis of storage and the changes of dynamic datasets over time, which is also known as the dynamic spatio-temporal graph problem. STINGER supports insertion and deletion of edges from scale-free graphs. Consequently, it allows fast query processing.  Another streaming partitioning algorithm was proposed by Tsourakakis \cite{Tsourakakis:2015:SGP:2817946.2817950} called the planted partition model. This model uses higher length walks for graph partitioning. As a result, it achieved negligible computational cost and it significantly improved the partition quality. Wang and Chiu proposed a scalable streaming partitioning approach, aiming to achieve a low complexity system. This partitioning technique\cite{Wang6691619} aims to reduce the edges between partitions, and thus reduce the communication cost for query processing. A workload streaming partitioning technique\cite{firth2016workload} was proposed to reduce inter-partition network communication. This technique is based on a well-known streaming graph partitioning heuristic\cite{Stanton:2012:SGP:2339530.2339722} that allocates vertices according to the maximum number of edges of a vertex in a partition. The technique overcame a few issues faced by previous algorithms, for example, inter-partition traversals when executing and pattern matching queries.  

A re-streaming algorithm \cite{Nishimura:2013:RGP:2487575.2487696} proposed by Nishimura et al. considered the scenario where the same datasets routinely streamed. This re-streaming technique performed well when same datasets repeatedly comes over in an application. However, drawback of this study is it is not suitable for the graphs where changes happen very frequently of the structure of the dataset. As such cases, data do not arrive in routine manner or don't repeat their stream. Consequently, re-streaming technique has less impact in such scenario.

A distributed vertex swapping technique called Ja-be-ja \cite{6676492} was proposed by Rahiman et al. this vertices swapping technique made uses to reduce the communication. Ja-be-ja was built based local search and Simulated Annealing(SA) method. SA method uses the statistical mechanism which is not suitable for the sparse network\cite{Williams:1991:PDL:124737.124739} 

Stanton\cite{Stanton:2012:SGP:2339530.2339722} proposed a few heuristics for partitioning a large-scale graph in a streaming manner. Linear Deterministic Greedy (LDG) was the best performing heuristic of those heuristics. This algorithm is a Greedy heuristic, which is linear. It has a central graph loader, which loads and distributes data among the available workers. The heuristic assigns a vertex to the partition with which it shares the most edges. The algorithm was evaluated using 21 different static datasets and up to 16 partitions. It makes heuristics scale with the size and number of graph partitions. Based on PageRank computations, the method yielded a significant speed up achievement for large social networks by 18\%-39\% when compared to Spark \cite{Zaharia:2010:SCC:1863103.1863113}. However, there are drawbacks in this study, which have addressed in our study. LDG receives the input data in a certain order, which is not suitable for any real-world streaming graph application whereas, WStream algorithm receives the graph input in sequentially as it arrives regardless of any particular order. Moreover, LDG algorithm uses the entire subgraph information by all vertices previously partitioned. And they used a distributed look up table to access the graph information. We addressed this issue in this study. Moreover, we used a stream window which creates more chances to get a better partitioning results by using a Greedy strategy.

The LDG algorithm is a well established streaming graph partitioning algorithm and is a state-of-the-art one-pass edge-cut partitioning algorithm. Therefore, in this study, we compared our one-pass edge-cut partitioning algorithm with the LDG algorithm \cite{Stanton:2012:SGP:2339530.2339722}.

\begin{table}[H]
	\centering
	\caption{Summarization of stream based graph partitioning}
	\label{related-work-table}
	\setlength{\tabcolsep}{3pt}
	\begin{adjustbox}{width=\linewidth} 
		\begin{tabular}{ |p{3cm}|p{1.5cm}|p{1.5cm}|p{1.5cm}|p{2.0cm}|  }
			\hline
			\multirow{2}{*}{Algorithm}{\textbf{}} & \multicolumn{4}{c|}{Features}{\textbf{}}\\ 
			\cline{2-5}
			& Vertex-cut & Edge-cut& Distributed & Window based\\
			\hline
			Linear Deterministic Greedy(LDG)\cite{Stanton:2012:SGP:2339530.2339722}  &  No   &Yes & Yes &No \\
			\hline
			Natural Graph Factorization\cite{Ahmed:2013:DLN:2488388.2488393}&Yes     &No &Yes  &No \\
			\hline
			LOOM\cite{firth2016workload} &No     &Yes &Yes  &No \\
			\hline
			STINGER\cite{Riedy:2013:MSD:2425676.2425689}&	No&	Yes&	No&	No\\
			\hline
			Planted Partition\cite{Tsourakakis:2015:SGP:2817946.2817950}&	No&	Yes&	No&	No\\
			\hline
			HDRF\cite{Petroni:2015:HSP:2806416.2806424}&	Yes&	No&	Yes&	No\\
			\hline
			HoVerCut\cite{7584914}&	Yes&	No&	Yes&	Yes\\
			\hline
			Vertex Migration\cite{Abdolrashidi7584916}&No&Yes&Yes&No\\
			\hline
		\end{tabular}
	\end{adjustbox}
\end{table}

We propose a window-based streaming processing algorithm; the window can contain more information about a candidate vertex that is ready to be assigned to a particular partition. While partitioning graph data from a window, the first vertex and its adjacent vertices of the window is assigned to a partition by the algorithm. The first entry checks for the presence of any connected vertices in the current window as well as the vertex with the most edges from the partitioned data. This stream window helps in deciding an appropriate partition to assign for the first entry and its associated vertices to a respective partition of the candidate vertex with the most edges or any of the connected vertices to reduce communication. This technique achieves significant improvement in reducing edge-cut because the decision to assign a vertex has some impact on its future connected vertex from the window. To the best of our knowledge, this technique has not yet been applied in any studies.

\section{Proposed Algorithm}

\subsection{Preliminaries}

We consider an undirected graph $G$, with a set of edges $ E$ and vertices $V$, such that $G=(V, E)$. A balanced k-way partitioning divides the graph into almost equal subsets. The graph partitioning algorithm uses a balancing constraint to keep all the partitions balanced. The balancing constraint can be defined by Equation (1):

\begin{equation}
\forall_i \in \{1..k\} : |V_i| \leq L_{max} := (1 + \alpha) \lceil |V|/k\rceil
\label{preliminary_eqn}
\end{equation}
Where, $\alpha$ is the imbalanced parameter, and is a non-negative real number. The vertex $v$ is adjacent to vertex $ u $ given there is an edge $ \{u,v\} \in E$. If vertex $v$ and vertex $ u $ reside in different partitions, this is called the \textit{cut edge}. Thus, $ E_{ij}:= \{\{u,v\} \in E : u\in V_i, v \in V_j \}$ is the set of edge-cuts between partitions. Edge-cut graph partitioning always aims at reducing this cut.
\subsection{System Architecture}
In our system, we used master machines, which is responsible for reading input, and assign the vertices to the clients. As such partitioning algorithm resides in the master machines. We used a Stream Generator which creates the stream data after receiving the input and then form a stream window. We used a distributed meta data file which has been used to store the information of vertices are already seen in the stream. This information was used to assign the future vertices from stream of data. Figure 1 shows the architecture of the system. 
\begin{itemize}
	\item \textbf{Master Machine:} Master machine receives the input graph data from the input file. In master machine the Stream Generator, generates the stream data and maintained the stream window before assigning each vertex to the respective partition. It also stores the partitioned vertex information and later this uses for future vertex partition. 
	\item \textbf{Partition:} Each partition also known as worker machine communicate with master machine to receive the assigned vertex from the master machine. Worker machine also communicates with each other to maintain the computation of a domain application.  
\end{itemize}
\subsection{The Streaming Model and Window}
We consider that the graph data comes in a stream of tuples $V<vertex;edges>$. The proposed algorithm utilizes a sliding stream window of size $W$. We define two different vertices in the stream window: 1) \textbf{\textit{Candidate vertex}} is the one, which is the front of the stream window and available for partitioning. 2) Neighbors of the candidate vertex in the window are known as a \textbf{\textit{Buffered vertex}} in this study. As shown in Figure 2, $V1$ is the candidate vertex that resides at the front of window which contains three more vertices and vertices $V3$ and $V4 $ are the neighbor of candidate vertex $V1$. Thus, these two vertices are defined as a buffered vertex in a stream window. 

\begin{figure}[H]
	\includegraphics[width=\linewidth]{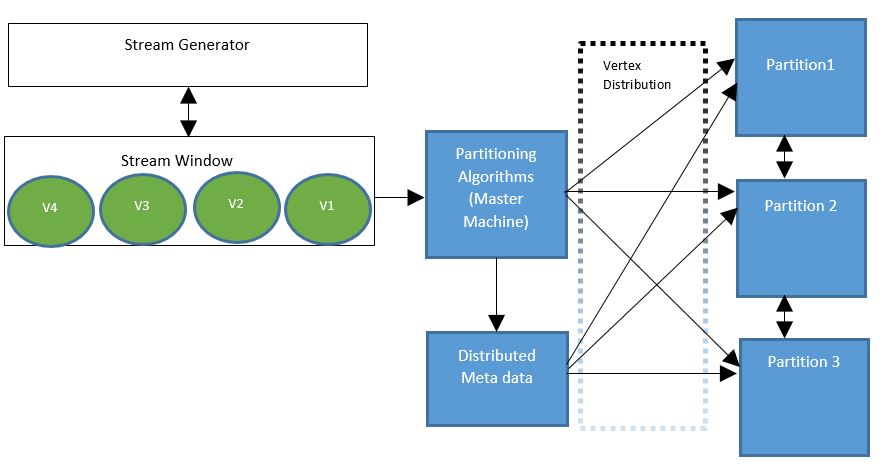}
	\caption{System Architecture}
	\label{fig:arch}
\end{figure}

\begin{figure}[H]
	\includegraphics[width=\linewidth]{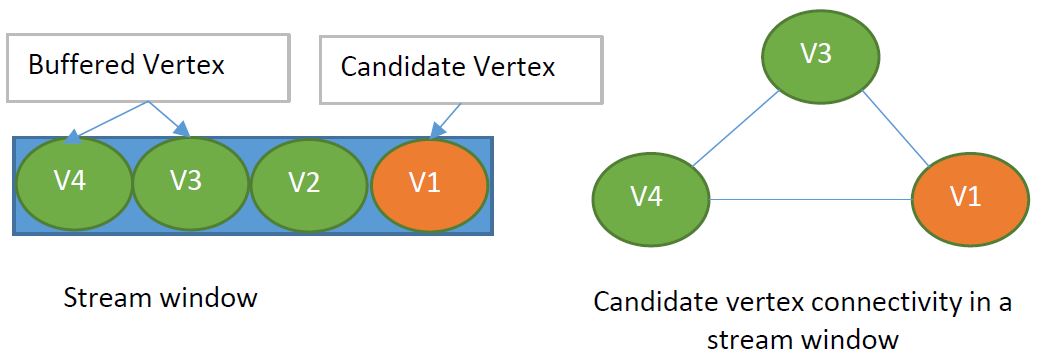}
	\caption{Candidate and Buffered Vertex in a Stream Window}
	\label{fig:window and connectivity}
\end{figure}

When a candidate vertex is allocated to a partition, it leaves a space for another vertex to come into the stream window to maintain the window size. The stream window contains more than one vertex such that it gives more information about a candidate vertex and other connected vertices of the candidate vertex. Consequently, this window-based partitioning produces better partitioning quality. The size of the window is $ W > 1$, and the size of the stream window depends on the graph structure and the type of graph. In this study, the minimum and maximum window size are 100 and 800, respectively.  

Stream order is another aspect to consider while performing streaming graph partitioning, as it has a major influence on the performance of graph partitioning. The input order of a graph makes a significant difference on the performance of a partitioning method. In a real-world graph with streaming settings, the order of a stream is not predictable. In this study, we consider a uniformly random order while receiving the graph input to the stream window. Algorithm 1 presents the pseudocode for our proposed WStream algorithm.

\begin{algorithm}
	\begin{algorithmic}
		\IF {all the partitions are empty}
		\STATE   assign $V$ randomly(uniform)  
		\ELSE
		\IF {$loadImbalance\geq B$}
		\STATE  perform greedy strategy except for the partition with the highest load
		\ELSE
		\STATE perform greedy strategy
		\ENDIF
		\ENDIF
	\end{algorithmic}
	\caption{WStream Algorithm}
\end{algorithm}

\begin{algorithm}
	\begin{algorithmic}
		\FOR{$i=0$ to $k$}		
		\IF {($|P_k \cap E(V_c)| \geq |P_k\cap E(V_b)|$)}
		\STATE $k\gets V_c$
		\ELSE
		\IF {($|P_k \cap E(V_c)| == |P_k\cap E(V_b)|$)}
		\STATE  $V_c$ to a partition randomly  
		\ELSE
		\STATE  $V_c$ to a partition that has minimum load
		\ENDIF
		\ENDIF	
		\ENDFOR 
	\end{algorithmic}
	\caption{Greedy Strategy}
\end{algorithm}
\subsection{WStream Algorithm}
The algorithm starts with three inputs, a number of vertices, their associated edges, and a balancing parameter. We also specified the number of partitions. The algorithm finds the total number of vertices of each partition and identifies the partition with the maximum number of vertices. The algorithm must keep track of differences among partitions to keep them balanced. Algorithm 1 shows the pseudocode for the WStream algorithm.

The balancing parameter $p$ checks the level of imbalance of the partitions with each other. We used the parameter p=[0, $\alpha$] and a range of $\alpha $ values to check the balancing performance of the algorithm. We observed that a higher  $\alpha$ value reduces the number of edge-cuts, and the load difference among partitions did not exceed the parameter $\alpha$. In the case of $\alpha$=0 the partitions were perfectly balanced. The balancing technique checks the load of the partitions after assigning a vertex to a partition. This is obtained by finding the load differences between partitions. Finding load differences means finding the comparison between the total numbers of allocated vertices among all the partitions.

During partitioning, if the difference between a partition with maximum load and any of the partitions exceeds the value of $\alpha$, the algorithm decides to assign vertices to other partitions using Greedy Strategy except for the partition with maximum load. After completing the partitioning task, we calculate the load imbalance by calculating the standard deviation of the number of vertices in each partition.  

This algorithm checks the balancing parameter after assigning each vertex to any partition. While performing the partitioning task, whichever partition exceeds the value of the parameter, the partitioning algorithm stop sending vertices to that partition and apply a Greedy strategy (discussed in the next section) to assign the following vertices to other partitions. The following section discusses in detail the mechanics of the vertex assignation technique using the Greedy strategy for partition decisions. 

\subsection{Greedy Strategy}

WStream algorithm exploit the Greedy Strategy to find the partition, which has the most edges of the candidate vertex or the buffered vertex from the window. Master machine is responsible for the partitioning task and distributes vertices to the clients. Master machine also stores the summary of vertex information to be used for assigning future vertices to an appropriate partition. At the beginning of partitioning task, the algorithm assigns the candidate vertex to a partition by using a uniform random distribution. Algorithm 2 shows the pseudocode of this window-based Greedy technique.

The major aim of this method is to find a partition for the candidate vertex and buffered vertex in the window. To decide this, neighbors of the candidate vertex (based on the graph summary) are also taken into account. The algorithm assign the candidate vertex along with its neighbors to the most weighted partition. This technique helps minimize the communication between partitions. Greedy technique tends to assign the vertices where they or their associated connections have the most connections. However, if the algorithm finds the same number of edges from two or more partitions the algorithm assigns that candidate vertex and its connected vertex from the window to the partition, which has fewer loads among the tied partitions. In any case, if the algorithm does not find any edges for candidate vertices and buffer vertices from a partition, it decide to assign the candidate vertex randomly in uniform manner to any of the partitions.

\begin{equation}
\underset{k \in P}{\operatorname{arg\,max}} \{|E(V_c) \cap P_k| \}
\label{eqn:02}
\end{equation}
\begin{equation}
\underset{k \in P, b\in B}{\operatorname{arg\,max}} \{|E(V_b) \cap P_k| \}
\label{eqn:03}
\end{equation}

Equations (2) and (3) show the formula to determine the partition, that has the maximum number of edges of the candidate vertex and buffered vertex, respectively. $ E(V_c)$ is the number of edges of the candidate vertex, $E(V_b)$ is the number of edges of the buffered vertex, $P_k$ is the set of vertices of the $kth$ partition, and $b\in B$ is the set of buffered vertex of the stream window.

\section{Performance Evaluation}

This section discusses the evaluation criteria, dataset, performance metrics and experimental environment used in this study. 

\subsection{Experimental Settings} 

We implemented our proposed WStream algorithm and the LDG algorithm\cite{Stanton:2012:SGP:2339530.2339722} by using JAVA programming language. We then compared these two algorithms by using two synthetic and five real-world graph datasets. Three performance metrics are used in this comparison. We also use the METIS graph partitioning algorithm to compare the partitioning performance. We consider different experimental scenarios to evaluate the WStream algorithm. We run our experiments on the Linux operating system using virtual machines from the Nectar cloud\cite{NectarCl23:online} service. The device has 12 GB of RAM and 4 VCPUs. 

\subsection{Dataset}
We evaluated our partitioning performance using several static undirected real and synthetic graph datasets from different graph data archives. Table 2 summarizes the basic characteristics of the datasets used in our experiments. We chose different sizes and a variety of graphs to observe the partitioning performance of the algorithm in the context of scalability. Different structure and volumes of data make differences in partitioning behavior and performance. For example, the degree of adjacent vertices of social networks data is positively correlated than another dataset\cite{Newman2003WhySN}.    

\begin{table}[H]
	\centering
	\caption{Dataset}
	\label{dataset-table}
	\setlength{\tabcolsep}{3pt}
	\begin{adjustbox}{width=\linewidth} 
		\large
		\begin{tabular}{ |p{2.5cm}|p{1.5cm}|p{3cm}|p{3cm}|p{1.5cm}|  }
			\hline
			\multicolumn{5}{|c|}{Characteristics of dataset } \\
			\hline
			\textbf{Name of Dataset}  &\textbf{$|V|$} &\textbf{$|E|$} & \textbf{Type} & \textbf{Source}\\
			\hline
			3elt (Synthetic)&4200 &13722 &Finite-element mashes &\cite{Chris}\\ 
			\hline
			GrQc&5242 &14496 &Collaboration Network &\cite{snapnets}\\ 
			\hline
			Wiki-vote&	7,115&	99,291&	Social&	\cite{snapnets}\\	
			\hline
			4elt (Synthetic)&	15,606&	45,878&	Finite-element mashes& \cite{Chris}\\	
			\hline
			AstroPh &	18,772	&198,110	&Citation&	\cite{snapnets}\\	
			\hline
			Email-enron&	36,692&	183,831&	Communication&	\cite{snapnets}\\
			\hline
			Twitter&	81,306&1,768,149& Social	&\cite{snapnets}\\
			\hline
			com-DBLP&	317,080&1,049,866& Citation	&\cite{snapnets}\\
			\hline
		\end{tabular}
	\end{adjustbox}
\end{table}

\subsection{Performance Metrics}
We observed the performance of our proposed algorithm using the following performance metrics: i) fraction of edge-cut; ii) load imbalance; and iii) execution time.
We observed the number of external connections of a vertex from one partition to another partition. We calculated the fraction of the edge-cut using Equation (4):

\begin{equation}
edge cut ratio =  \frac{|E(u,v)|}{|E|} 
\label{eqn:04}
\end{equation}
Where, $|E| $ is the total number of edges of a graph and $ |E(u,v)|$ the total number of edges between $u$ and $v$ across partitions.
We calculated the standard deviation of the number of vertices in a partition to observe the imbalance from one partition to another. Equation (5) was used to calculate the load imbalance: 
\begin{equation}
load Imbalance=   \sqrt  \frac{\Sigma|v-\bar{v}|^2}{n} 
\label{eqn:05}
\end{equation}

Where, $v$ is the total number of vertices of a partition and $n$ is the total number of partitions. 

We measured the execution time from the start of partitioning until the end of partitioning. Input receiving time is also calculated during this execution time, as the streaming partitioning algorithm executes as stream data arrives.  

\subsection{Evaluation Scenario}
We used an unrestricted stream model to receive our graph input. In this model, the algorithm accepts a static graph input in sequential order and in a one-pass manner. We used a different combination of partitions (e.g. 2, 4, 8, and 16) with different ranges of datasets. 

To evaluate the effectiveness of our algorithm, we evaluated how the variation of stream window sizes affects performance; we conducted experiments with the following window sizes: 100, 200, 300, 400, 500, 600, 700 and 800. The size of a window refers to the number of vertices in a window with its associated edges. Figure 1 depicts the impact of window sizes on the WStream algorithm performance.

Variations in the balancing factor with different balancing parameters were also evaluated in this study. We conducted experiments with different balancing parameters such as 50, 100, and 150. 

In this evaluation, we also compared the performance of our algorithm with the best streaming graph partitioning heuristic currently in the field, the LDG algorithm \cite{Stanton:2012:SGP:2339530.2339722}. This algorithm employs a state-of-the-art streaming graph partitioning technique. A few studies have been done on streaming graph partitioning, however, most have not used the one-pass streaming in their implementation. The LDG algorithm is similar to our method and it is also one of the best streaming graph partitioning techniques in this field. This is the main reason for the selection of this algorithm for comparison purposes. 

METIS is one of the most powerful algorithms for offline graph partitioning and it is a near-optimal algorithm. Therefore, we also compared our WStream algorithm with METIS to observe the edge-cut performance. That gives us an insight as to how close WStream is to an optimal algorithm regarding minimizing the number of edge-cuts.     

\section{Results Analysis}
This section discusses the results of the different datasets using different scenarios such as a different number of stream windows, a different number of partitions and different balancing parameters. We also compare our evaluated results with a state-of-the-art algorithm \cite{Stanton:2012:SGP:2339530.2339722} for streaming graph partitioning. 

\subsection{Impact of Stream Window Sizes}

The WStream algorithm generates a stream window while receiving a stream of graph data. In this evaluation, we use a different number of window sizes to observe the partitioning performance (e.g. a reduction in communication cost) of the algorithm.

\begin{figure*}
	\label{Impact-of-window-sizes}
	\centering
	\subfloat[3elt Dataset]{%
		\includegraphics[width=.5\textwidth]{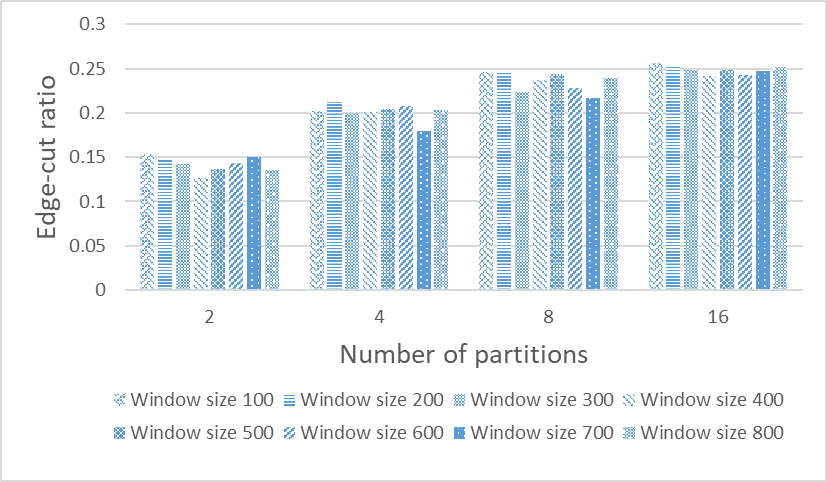}}\hfill
	\subfloat[4elt Dataset]{%
		\includegraphics[width=.5\textwidth]{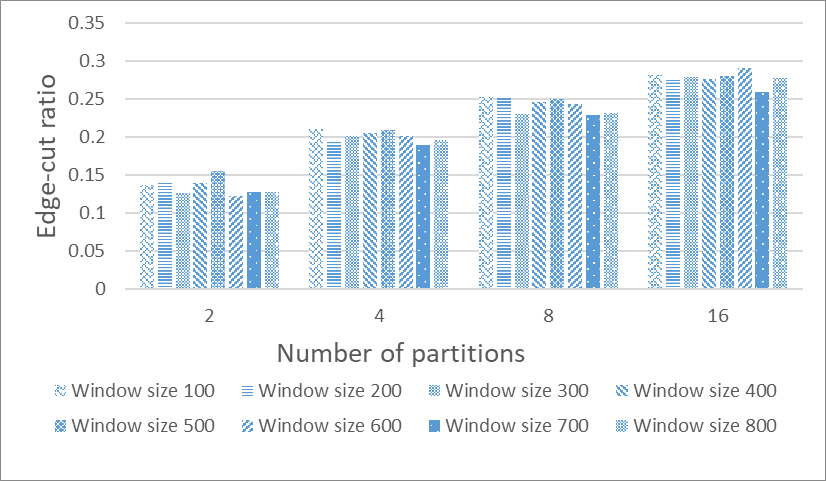}}\hfill
	\subfloat[AstroPh Dataset]{%
		\includegraphics[width=.5\textwidth]{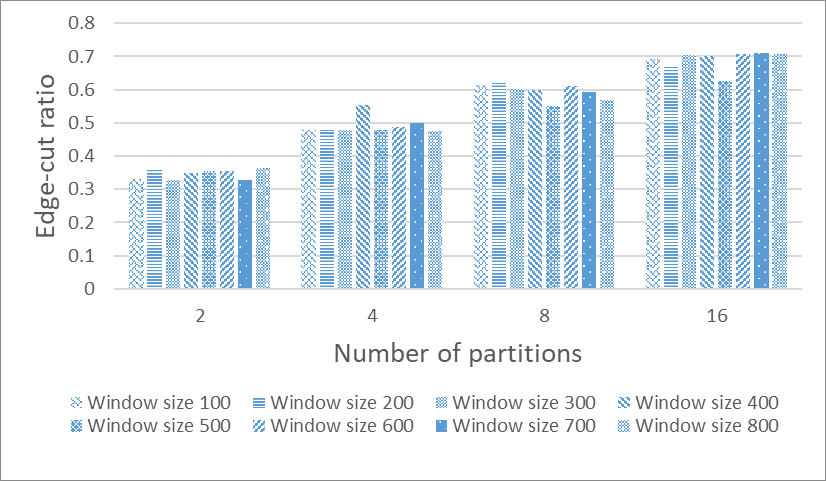}}\hfill
	\subfloat[GrQc Dataset]{%
		\includegraphics[width=.5\textwidth]{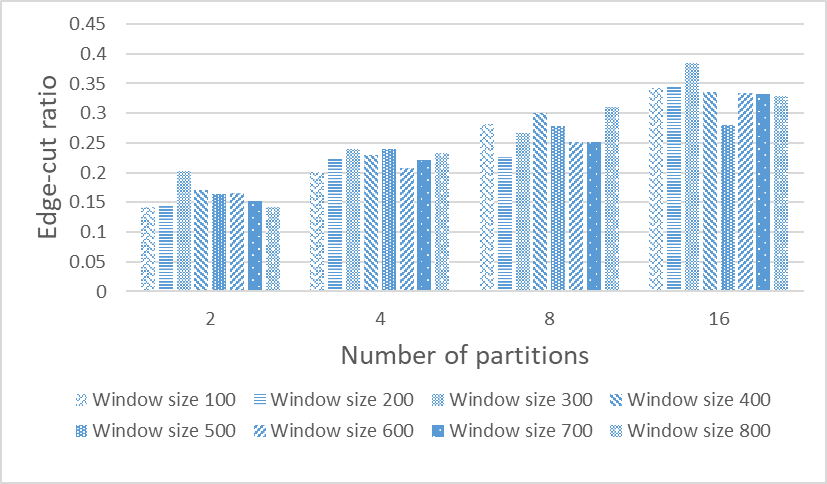}}\hfill
	\subfloat[Email-enron Dataset]{%
		\includegraphics[width=.5\textwidth]{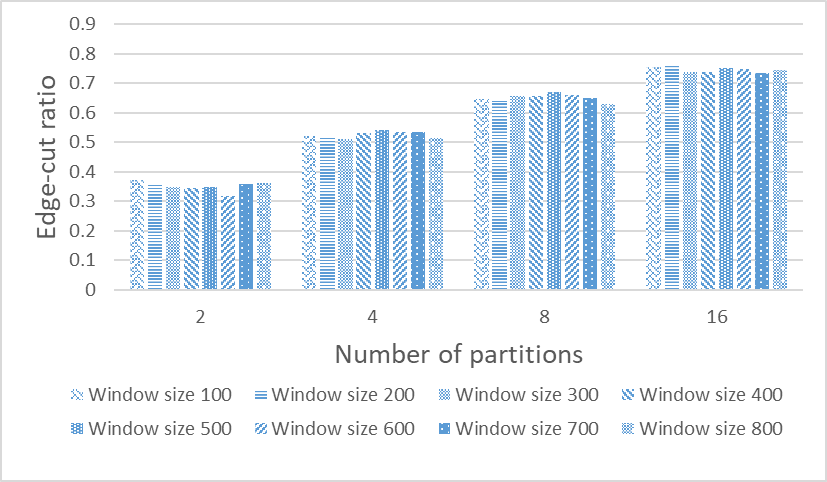}}\hfill
	\subfloat[Wiki-vote Dataset]{%
		\includegraphics[width=.5\textwidth]{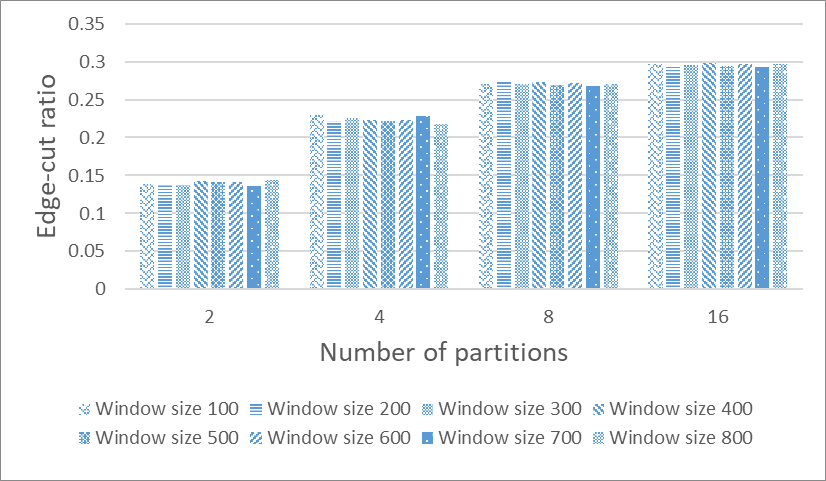}}\hfill
	\subfloat[Twitter Dataset]{%
		\includegraphics[width=.5\textwidth]{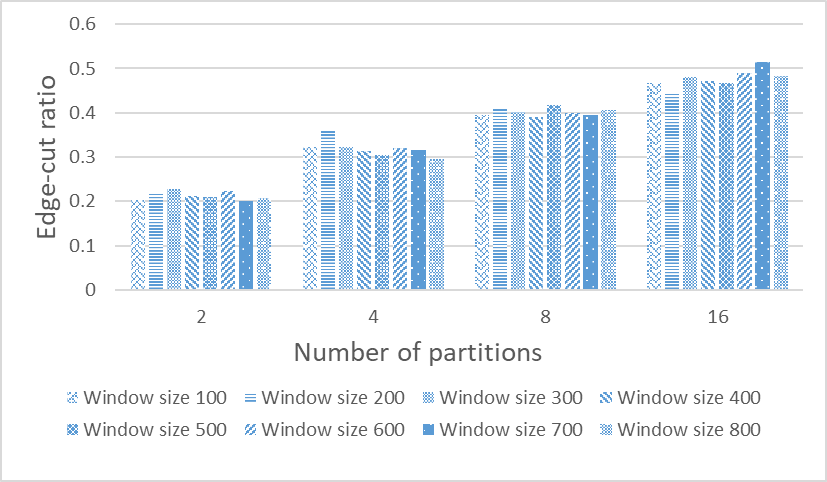}}\hfill
	\subfloat[com-DBLP Dataset]{%
		\includegraphics[width=.5\textwidth]{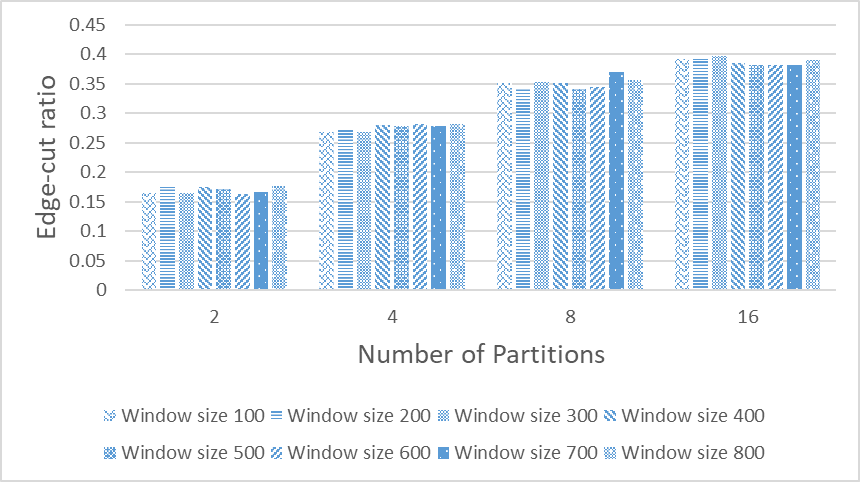}}\hfill
	\caption{Impact of different window sizes on different datasets}
\end{figure*}
Figure 3 depicts the edge-cut ratio for different partitions with varying window sizes. The WStream algorithm aims to utilize the stream window to provide efficient partitioning performance. The results show the impact of different window sizes on the partitioning performance of WStream algorithm. It is expected that the edge-cut decreases with a larger window size, as a larger window contains more information of a candidate vertex. However, as shown in Figure 3, in some cases, a higher edge-cut ratio is obtained for the larger stream window sizes compared to the smaller window sizes. This is because our balancing parameter checks the imbalance of all the partitions every time stream data is received before allocating them to a partition. To keep the partitions balanced according to our balancing parameter, the algorithm defies the Greedy strategy and allocates the vertices to the partition with the minimum load. In this case, the WStream algorithm does not obtain the expected edge-cut reduction.

Figure 3(g) demonstrates the edge-cut ratio of the Twitter dataset with a different number of stream window sizes and different partitions. We observe that the 4 partitions setting performs well and achieves expected outcome for the window-based streaming algorithm, except for window size 200 and window size 600, which obtains a slightly higher edge-cut, otherwise it reduces the edge-cut as we increase the window size.

Figure 3(e) shows that the WStream algorithm performs well for the Email-enron dataset for 2 partitions and 16 partitions in reducing number of edge-cuts as the window size increases.

As depicted in Figure 3(g) and 3(h) for com-DBLP and Twitter dataset which contains billion edges. WStream performs well in reducing the edge-cut as the window size increases. We observe that, for the 16 partition com-DBLP dataset gets better performance than other partition settings and Twitter datasets performs well for the 2 partition settings.   

\subsection{Impact of Balancing Factor}
Figure 4 shows the impact of the balancing parameter in reducing the communication cost of our WStream algorithm for the Twitter dataset. It can clearly be seen that, as expected, a larger balancing parameter reduces the inter-partition communication across partitions, because the smaller balancing parameter provides lesser load imbalance. To make partitions more balanced in a cluster, vertices have to be distributed to the machines; thus resulting in increased communication cost. In the WStream algorithm, the applied balancing parameter provides greater reduction in communication cost. The more partitions are balanced the more communication it creates. As depicted in Figure 4, the balancing parameter at the value of 50 produces about 17\%- 19\% more edge-cuts compared to the value of 150 for all partition settings. The proposed WStream algorithm checks the workload of each partition prior to assigning a vertex to a partition, where the difference in workload among partitions never exceeds the value of balancing parameter. Consequently, it keeps the imbalance as low as possible across partitions.

Figure 5 shows the load imbalance in the Twitter dataset with different balancing parameters for partitions 2, 4, 8, and 16. It is clearly seen that the smaller balancing parameter has a lower imbalance for any number of partition settings.

\subsection{Performance Comparison}
\subsubsection{Edge-cut comparison}
Figure 6 shows the comparison of the fraction of edge-cut between the LDG, METIS and the WStream algorithms for different scales of datasets. Our algorithm outperforms the LDG algorithm in reducing the edge-cut for all datasets except for the Wiki-Vote dataset. Edge-cut ratio differences between METIS and WStream is quite promising as METIS is a static and near optimal graph partitioning algorithm.

In this evaluation, we use a different number of partitions (e.g. 2, 4, 8, and 16) to test the partitioning performance. As expected, the communication cost increased for a larger number of partitions, as shown in Figure 6, for all the algorithms. We compare the edge cut of different datasets. In this comparison, we use a window size of 100 for the WStream algorithm. A window size of 100 provides the worst performance of the WSteam algorithm as expected. That is why, we choose this least performance of WStream algorithm to make comparison with LDG algorithm. From Figure 6, it is clearly seen that the WStream algorithm outperforms the LDG algorithm in reducing the edge-cut for the 3elt dataset. The result indicates that the WStream algorithm is able to reduce the edge-cut ratio by 56\% for 16 partitions for the 3elt dataset. As demonstrated in Figure 6(a), our WStream algorithm shows significant improvement in reducing edge-cuts by about 40\%-56\% for all partition settings. WStream algorithm has significant improvement in reducing edge-cut ratio compared to LDG algorithm for the large scale dataset like com-DBLP and Twitter. It is noticed that, it reduces 66\%-75\% edge-cut ratio for these two datasets.  
\begin{figure}[H]
	\includegraphics[width=\linewidth]{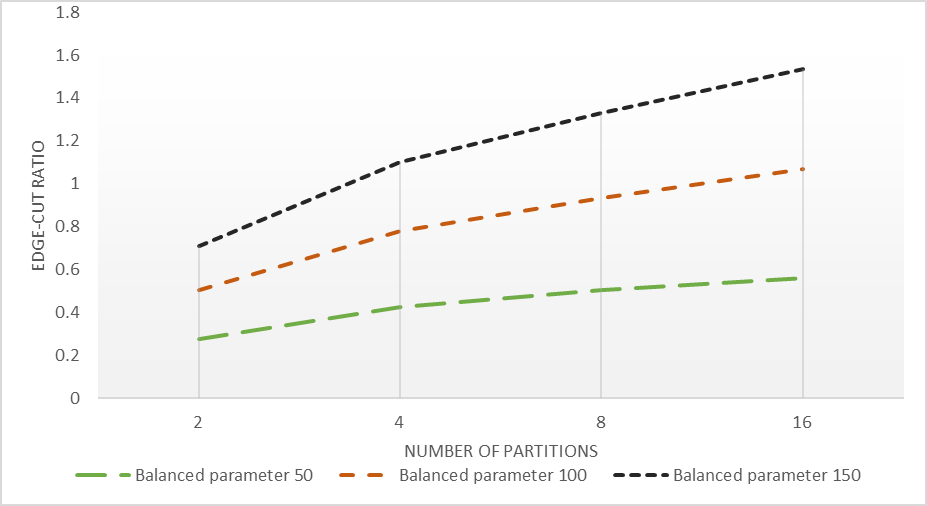}
	\caption{Variation of Balancing Parameter for Twitter dataset}
	\label{varius_parameter_fig}
\end{figure}

\begin{figure}[H]
	\includegraphics[width=\linewidth]{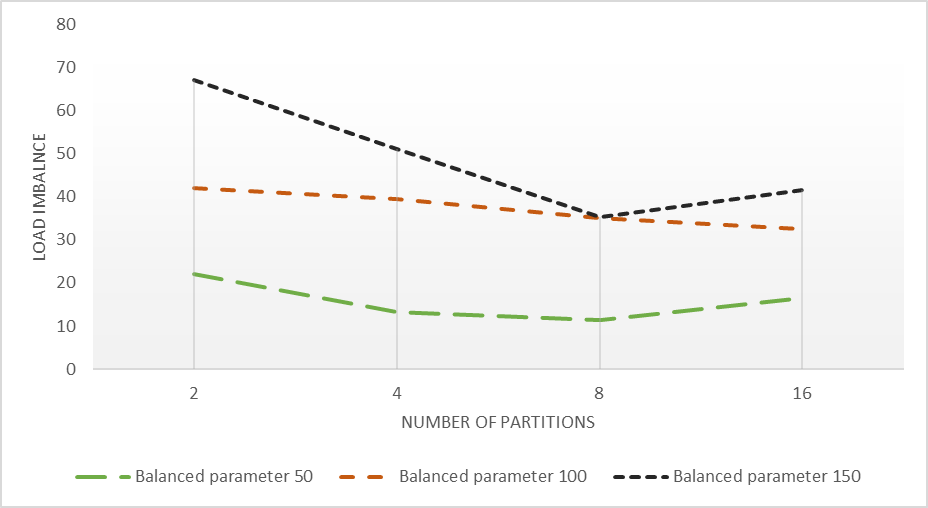}
	\caption{Load Imbalance for Twitter dataset}
	\label{load_imbalance_fig}
\end{figure}

\begin{figure*}
	\label{edge-cut-comparison}
	\centering 
	\subfloat[3elt Dataset]{%
		\includegraphics[width=.5\textwidth]{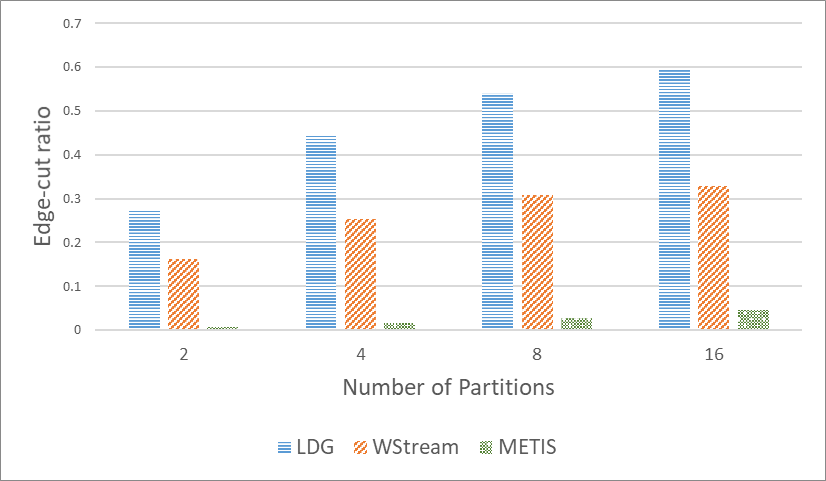}}\hfill
	\subfloat[4elt Dataset]{%
		\includegraphics[width=.5\textwidth]{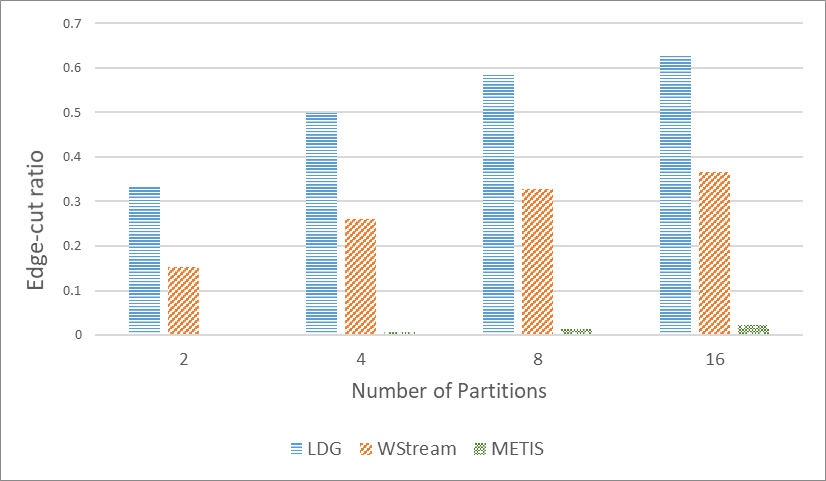}}\hfill
	\subfloat[AstroPh Dataset]{%
		\includegraphics[width=.5\textwidth]{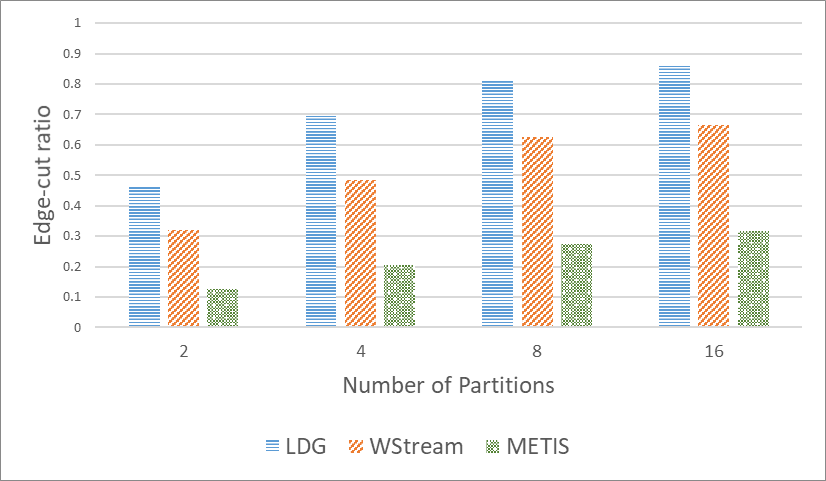}}\hfill
	\subfloat[GrQc Dataset]{%
		\includegraphics[width=.5\textwidth]{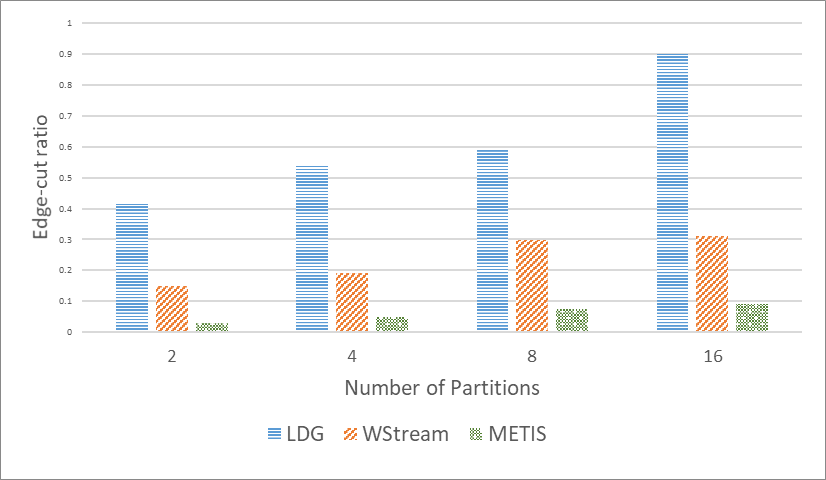}}\hfill
	\subfloat[Email-enron Dataset]{%
		\includegraphics[width=.5\textwidth]{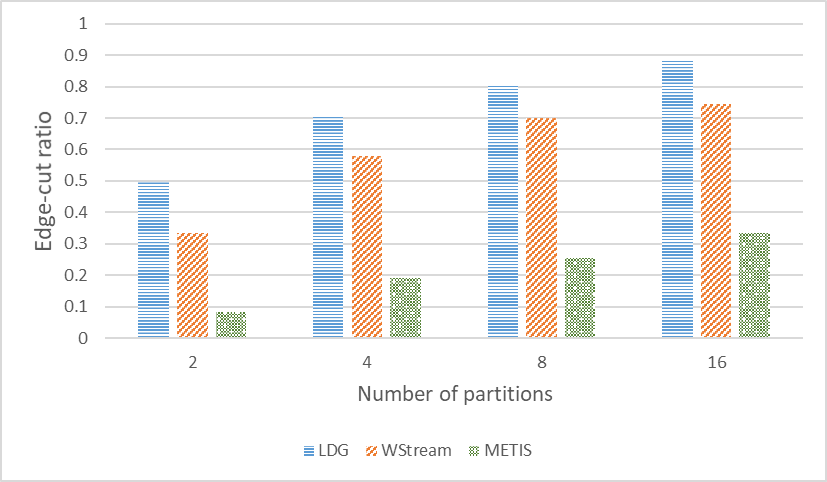}}\hfill
	\subfloat[Wiki-vote Dataset]{%
		\includegraphics[width=.5\textwidth]{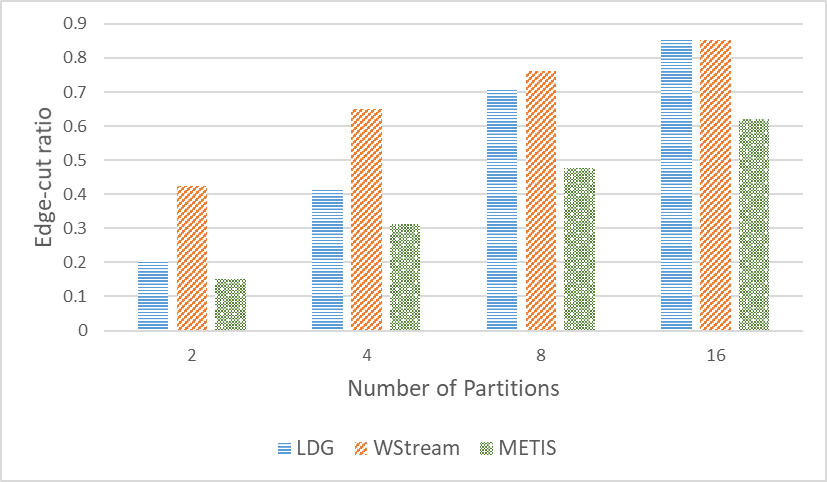}}\hfill
	\subfloat[Twitter Dataset]{%
		\includegraphics[width=.5\textwidth]{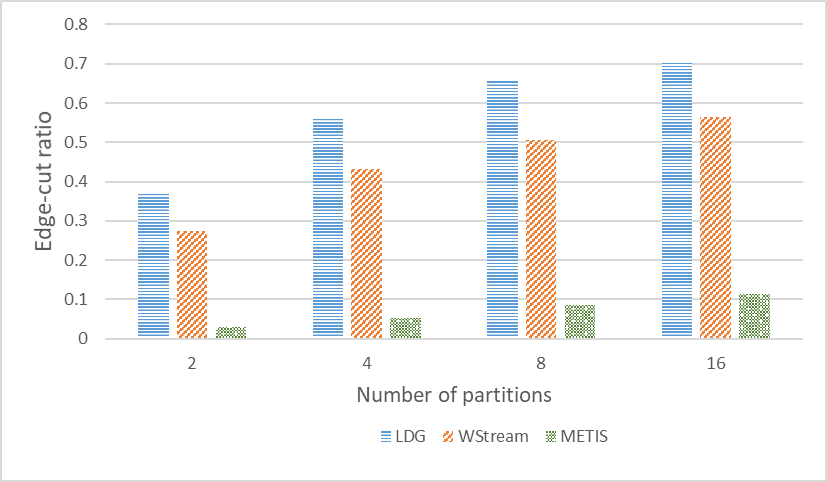}}\hfill
	\subfloat[com-DBLP Dataset]{%
		\includegraphics[width=.5\textwidth]{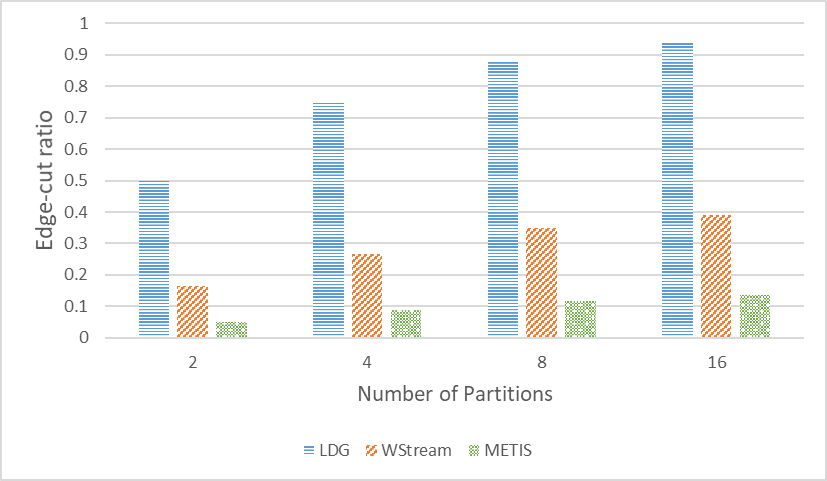}}\hfill
	\caption{Edge-cut ratio comparison}
\end{figure*}


Our WStream algorithm is compatible with datasets of different scales and outperformed the LDG algorithm in most cases. However, there is slight performance degradation of WStream algorithm for Wiki-Vote dataset. Figure 6(f) depicts the edge-cut comparison between the WStream and LDG algorithms for the Wiki-Vote dataset. The results indicate that the performance of the WStream algorithm drops slightly compared to the LDG algorithm for this dataset. This is due to the behavior of the dataset. This dataset is structured with a higher degree of vertex distribution compared to other datasets in this study. While partitioning a graph, the algorithm tends to make partitions as balanced as possible according to the balancing parameter. Consequently, the adjacent vertices of a vertex might have been allocated to other partitions. Thus, this causes more edge-cuts in the Wiki-Vote dataset.

Figure 6 also depicts the edge-cut performance of the METIS algorithm along with WStream and LDG algorithms. As expected, METIS outperforms LDG and WStream algorithm. Because, it is an offline graph-partitioning algorithm where all the information about the graph is known is prior to partitioning. METIS provides optimal partitioning but for the AstroPh, Email-enron and Wiki-vote datasets overall edge-cut difference is 20\%-25\% between METIS and WStream algorithms. This is quite promising performance of WStream algorithm when compared to a static graph partitioning algorithm. The worst performance was for WStream for the Twitter, 3elt, and 4elt dataset compared to METIS. 

In conclusion, our WStream algorithm was able to significantly reduce edge-cut for different graphs ranging from 3000 vertices to 317080 vertices.

\subsubsection{Load Imbalance Comparison}
We compare the load imbalance of our proposed algorithm with the LDG algorithm and METIS for every partition setting. Table 3 shows the load imbalance between the LDG algorithm, WStream and METIS for different datasets. We use the WStream algorithm partitioning result with a window size of 100 for this comparison.

The WStream algorithm achieves a completely balanced allocation for the 3elt dataset with the 2 partitions setting. The LDG algorithm achieves a completely balanced partitioning for most of the datasets except for the Wiki-Vote in the 2 partitions setting. However, for the 16 partitions, the WStream algorithm demonstrates an 85\% load imbalance reduction compared to the LDG algorithm for the 4elt and WikiVote datasets. For all partitions, the Wiki-vote was better balanced using the WStream algorithm compared to LDG algorithm, and load imbalance in the 3elt and GrQc datasets is significantly reduced. It is quite balanced for the AstroPh dataset for 16 partitions by the WSstream algorithm. However, WStream does not perform well for the com-DBLP, Twitter and Email-enron dataset in balancing the load for any partition number, but it is much more efficient in reducing communication than LDG algorithm. 

The LDG algorithm uses a capacity constraint to keep the partitions balanced. However, our WStream algorithm uses a balancing parameter to keep all the partitions balanced at a certain threshold. In reality, different real-world applications behave differently, where some graph partitioning applications require that the computation load be balanced. On the other hand, some applications require communications to be reduced across partitions. Based on this reality, the graph partitioning objective should be whether to minimize the load imbalance or minimize communication. This is very subjective and application-dependent. Our WStream algorithm performs well in general, and it is application independent. It also offers a trade-off between load imbalance and edge-cut minimization.

The WStream algorithm achieves load imbalance reduction for every dataset except for 3elt compared to METIS. WStream algorithm outperforms METIS for the big datasets such as com-DBLP, Twitter, Email-enron and AstroPh for every partition setting. WStream achieves 99\% reduction for AstroPh and Email-enron for 2 partitions setting and more than a 77\% load imbalance reduction for the AstroPh, Email-Enron, and Twitter datasets for every partition setting. It is clearly seen that, WStream achieves 97\%-99.5\% load imbalance reduction for the big dataset with billion edges like com-DBLP. However, except 2 partitions setting WStream performs better than METIS.         

\begin{table}[H]
	\centering
	\caption{Load Imbalance Comparison}
	\label{Load-Imbalance-table}
	\setlength{\tabcolsep}{3pt}
	\begin{adjustbox}{width=\linewidth}
		\large
		\begin{tabular}{|p{2cm}|p{1cm}|p{1.5cm}|p{1.2cm}|p{1cm}|p{1.5cm}|p{1.2cm}|p{1cm}|p{1.5cm}|p{1.2cm}|p{1cm}|p{1.5cm}|p{1.2cm}|} 
			
			\hline
			\multirow{4}{*}{Dataset} & \multicolumn{12}{c|}{Load Imbalance (standard deviation)}    \\ \cline{2-13} 
			& \multicolumn{12}{c|}{Number of Partitions}            \\ \cline{2-13} 
			& \multicolumn{3}{c|}{2} & \multicolumn{3}{c|}{4} & \multicolumn{3}{c|}{8} & \multicolumn{3}{c|}{16} \\ \cline{2-13} 
			& LDG & WStream &METIS &LDG & WStream &METIS & LDG & WStream &METIS& LDG &  WStream&METIS       \\ \hline
			3elt  & 0.0      & 7.0 & 4       &  19.01          & 17.54&6.04       &    56.71   & 7.63&4.69   & 22.62  & 7.56&3.16 \\ \hline
			GrQc &     0.0      &  25.0&74     & 258.94   & 18.95&22.01  & 324.20    & 15.16&14.79    &  65.94    & 9.96&6.44 \\ \hline
			4elt    &     2.5      &  5.5&2     & 3.42          & 6.98&9.2   & 0.33 &  14.27&32.32   & 94.23     &11.68&11.69  \\ \hline
			Wiki-Vote    &     213.5     &  24.5&100.5  &    285.93      &  18.95 &38.05&	210.85&15.93&21.32& 92.89&	11.73&13.22   \\ \hline
			AstroPh    &  0.0	&1.0&274&0&	9.72&137.03&	0.5	&14.19&66.84&	272.08&	8.86&30.88\\ \hline
			Email-Enron    & 0.0	&1.0&551&	0&	24.50&268.01&	0.5	& 19.46&137.28&	0.43	& 13.99&59.89 \\ \hline
			Twitter    &  0.0	&25.0&254&0.5&18.20&591.1	&0.43&15.56&252.74&0.48&15.29&115.45 \\ \hline
			com-DBLP    &  0.0	&22.0&4261&0.0&9.20&1531.16	&0.0&16.80&659.09&0.5&11.95&418.99 \\ \hline
		\end{tabular}
	\end{adjustbox}
\end{table}

\section{Conclusions and Future Direction}
In this study, we have studied streaming graph partitioning by edge-cut. Stream-based partitioning has become very prominent recently due to large-scale expansions of social media graphs, which require distributed processing. Several algorithms have been proposed to partition data in a stream and thus reduce the execution time for partitioning, while keeping load imbalance and edge-cut to a minimum.

In this paper, we demonstrated that a stream window in streaming graph partitioning results in significantly high quality partitioning. We proposed a streaming partitioning algorithm for large-scale graphs using a streaming window to minimize the edge-cut across partitions while reducing the imbalance among partitions. We compared our proposed WStream algorithm with the state-of-the-art LDG algorithm using real and synthetic graph data sets. The evaluation results clearly show that the WStream algorithm reduced the edge-cut by 40\%-56\% in comparison to the LDG algorithm for all datasets except for the Wiki-Vote dataset. In terms of load balancing, our algorithm performed better than the LDG algorithm for 16 partitions except for the Email-Enron and Twitter datasets, and performed extremely well for the 4 partitions and 8-partitions for most of the datasets. However, our proposed algorithm partitions the graphs faster. This makes the proposed algorithm quite suitable for cases where graphs are growing at a rapid speed. 
In the future, we will extend this study to include the evaluation of dynamic graphs.

\bibliographystyle{unsrt}
\bibliography{reference}
	
\end{document}